# Deep Learning-Based Partial Volume Correction in Standard and Low-Dose PET-CT Imaging

Mohammad-Saber Azimi[1], Alireza Kamali-Asl[1], Mohammad-Reza Ay[2], Navid Zeraatkar[3], and Hossein Arabi[4],

[1] Department of Medical Radiation Engineering, Shahid Beheshti University, Tehran, Iran

[2] Department of Medical Physics and Biomedical Engineering, Tehran University of Medical Sciences, Tehran, Iran

[3] Siemens Healthineers, Knoxville, TN, USA

[4] Division of Nuclear Medicine & Molecular Imaging, Geneva University Hospital, CH-1211, Geneva, Switzerland

**Abstract**

A standard dose of radioactive tracer must be delivered into the patient's body to obtain high-quality Positron Emission Tomography (PET) images for diagnostic purposes, which raises the risk of radiation harm. A reduced tracer dose, on the other hand, results in a poor image quality and a noise-induced quantitative bias in PET imaging. The partial volume effect (PVE), which is the result of PET intrinsic limited spatial resolution, is another source of quality and quantity degradation in PET imaging. The utilization of anatomical information for PVE correction (PVC) is not straightforward due to the internal organ motions, patient involuntary motions, and discrepancies in the appearance and size of the structures in anatomical and functional images. Furthermore, an additional MR imaging session is necessary for anatomical information, which may not be available. We set out to build a deep learning-based framework for predicting partial volume corrected full-dose (FD+PVC) pictures from either standard or low-dose (LD) PET images without requiring any anatomical data in order to provide a joint solution for PVC and denoise low-dose PET images. To this end, we trained a modified encoder-decoder U-Net network with standard or LD PET images as the input and FD+PVC images by six different PVC methods as the target. These six PVC approaches include geometric transfer matrix (GTM), multi-target correction (MTC), region-based voxel-wise correction (RBV), iterative Yang (IY), reblurred Van-Cittert (RVC), and Richardson-Lucy (RL). Different levels of error were observed for these PVC methods, which were relatively smaller for GTM with a structural similarity index (SSI) of 0.63 for LD and 0.29 for FD, IY with an SSI of 0.63 for LD and 0.67 for FD, RBV with an SSI of 0.57 for LD and 0.65 for FD, and RVC with an SSI of 0.89 for LD and 0.94 for FD PVC approaches. However, large quantitative errors were observed for multi-target MTC with a root mean squared error (RMSE) of 2.71 for LD and 2.45 for FD and RL with an RMSE of 5 for LD and 3.27 for FD PVC approaches. We found that the proposed framework could effectively perform joint denoising and PVC for PET images with LD and FD input PET data (LD vs. FD). When no MR images are available, the developed deep learning models could be used for PVC on LD or standard PET-CT scans as an image quality enhancement technique.

## I. Introduction

As a nuclear medicine imaging technology, positron emission tomography (PET) is the cornerstone of clinical practice for therapeutic evaluation and diagnostic purposes [1, 2]. A sufficient amount of radioactive tracer must be delivered into the patient's body to obtain a high-quality PET image, which introduces the risk of radiation damage [3-6]. The dosage of the injected radiotracer should be decreased to an acceptable level to limit the radiation risk, based on the well-known as low as reasonably achievable (ALARA ) principle [7]. A reduced radiotracer dosage would result in a poor PET image quality, as well as a noise-induced quantitative bias, considering the stochastic nature of signal formation in PET imaging [8-10]. Moreover, the partial volume effect (PVE), which is the result of PET inherent finite spatial resolution in PET imaging, causes large bias, especially for structures with comparable sizes to the point spread function (PSF) of the system, spill-in and spill-out across the neighboring regions, and ambiguity at tissue boundaries[7, 8, 11]. In this light, prior to the quantitative evaluation of metabolism and physiology of the organs/lesions, partial volume correction (PVC) and noise suppression are critical in PET studies.

The PVC techniques may be divided into two categories: reconstruction-based and post-reconstruction approaches [12-15]. In the reconstruction-based method, anatomical data could be incorporated into an iterative image reconstruction method, as prior information, to compensate for the PVE across different anatomical regions. This implies that the reconstruction algorithm is guided by an anatomical signal from a jointly acquired MR image [16]. On the other hand, post-reconstruction PVC is commonly used to estimate the correct uptake values for regions of interest using anatomical regions derived from structural imaging, such as MRI. Post-reconstruction PVC approaches tend to estimate correct uptake values either at the voxel level or anatomical (regions of interest) regions [17].

Regarding the conventional noise reduction techniques in emission tomography [18-21], machine learning approaches have demonstrated an outstanding ability to restore the image quality in emission tomography [6, 20, 22, 23]. In this regard, Xu et al. developed a fully convolutional encoder-decoder residual deep network model to estimate standard-dose PET images from ultra-low-dose data. Their proposed deep network outperformed the non-local mean (NLM) and block-matching 3D filters in terms of peak signal-to-noise ratio (PSNR), structural similarity index (SSI), and root mean squared error (RMSE) [24]. Cui et al. proposed an unsupervised deep learning strategy for PET denoising, in which the network was given the patient's previous scans, in addition to the noisy PET image, as the training label. The prior image of the patient, in addition to the initial noisy PET image, was supplied to the network as an instructional label. Their findings demonstrated that the proposed approach outperformed the conventional Gaussian, NLM, and block-matching 4D filters [25].

A number of approaches have been proposed for the correction of PVE in PET imaging (post-reconstruction methods), which include Geometric transfer matrix (GTM) [26], multi-target correction (MTC) [27], region-based voxel-wise correction (RBV) [28], iterative Yang (IY) [29], reblurred Van-Cittert (RVC) [30], and Richardson-Lucy (RL) [30]. The first four PVC methods are region-based and require anatomical information and/or regions of interest, while the latter two are deconvolution-based applied at the voxel level. Post-reconstruction PVC and denoising approaches are frequently employed in PET imaging; however, a joint estimate of the partial volume corrected and denoised PET images would be crucial particularly in low-dose (LD) PET imaging [31]. In this regard, Xu et al. suggested a framework for a combined PET image denoising, PVC, and segmentation by an analytical approach. Their findings demonstrated that the proposed framework could effectively eliminate noise and correct PVE [32]. For joint PVC and noise reduction, Boussion et al. integrated deconvolution-based approaches (RL and RVC) with wavelet-based denoising. This method improved the accuracy of tissue delineation and quantification without sacrificing functional information when tested on clinical and simulated PET images [31].

Since most PVC approaches rely on anatomical information obtained from MRI scans [33], co-registration of PET and MRI data is required. The utilization of anatomical information in PVE correction would not be straightforward due to the absence of MR images in most clinical routines, internal organ motion, patient uncontrolled movements, and changes in the appearance and size of structures in anatomical and functional imaging [31, 34]. In this study, we proposed a deep learning solution for joint PVC and noise reduction in LD PET imaging. Considering LD and full-dose (FD) PET images as the input (separately) for the deep learning solution, six frequently used PVC techniques were applied to FD PET images (FD+PVC) to serve as the targets for the training of the deep learning model. A modified encoder-decoder U-Net was trained using either LD or FD PET images as the input and FD+PVC (by six different methods) as the reference. The proposed deep learning solution would perform joint noise reduction and PVC on LD PET images.

## II. Materials and Methods

### A. Data acquisition

Clinical $^{18}$F-FDG brain PET/CT images (head and neck cancer) were collected for 20 min on a Biograph 16 PET/CT scanner (Siemens Healthcare, Germany) following a standard injection dose of 205 10 MBq for 160 patients (100 subjects for training, 20 subjects for validation, and 40 subjects for test). The PET data were obtained in a list-mode format to allow for the generation of synthetic LD data by randomly selecting 5% of the events within the acquisition time. For deep learning training, the reconstructed PET images were transformed into a matrix of 144×144×120 with a voxel size of 2 mm.

### B. Partial volume correction approaches

The PETPVC toolbox [33] was used to execute PVC on the PET images of 160 patients. The PETPVC was created utilizing the Insight Segmentation and Registration Toolkit (Insight Software Consortium, USA) in a C++ environment. Six distinct PVC techniques were used in this study, including GTM [26], MTC [27], RBV [28], IY [29], RVC [30], and RL [30]. We employed the automated anatomical labeling (AAL) [35] brain atlas transferred/co-registered to the patients' PET scan to apply PVC. The AAL brain mask was employed for four of these PVC techniques, namely GTM, MTC, RBV, and IY, since these techniques are region-based and require anatomical masks for implementing PVC. The brain regions mask was not used for the two deconvolution-based algorithms, namely RVC and RL. The six PVC methods are briefly discussed in the following sections.

*B.1. Geometric transfer matrix (GTM):*

The GTM [26] is a region-based method for PVC, which depends on anatomical information in terms of boundaries between different regions. The GTM is formulated by Eq. (1)

$$T(N, 1) = G^{-1}(N, N) \times r(N, 1) \tag{1}$$

Wherein *T* represents real activity uptakes in each anatomical (brain) area, *G* represents the spill-over of activity from one region to another, registered in a matrix of size *N* (numbers of regions), and *r* is the vector of mean activity in each region as measured from the input (original) PET data. The spill-over activity ($G_{ij}$) of the region *(i)* into the region *(j)* is calculated through the smoothing area *(i)* with the system's point spread function and multiplying with the region *(j)*. The remaining voxels are then added together and normalized by the total activity in the area *(i)*. The GTM method assumes that activity within an area is uniform, and a single value would represent its activity uptake.

*B.2. Multi-target correction (MTC):*

The MTC [27] is a combined region- and voxel-based PVC technique, wherein the mean activity levels in each region are first determined using the GTM approach, and thereafter, the whole area is corrected voxel-by-voxel. The MTC PVC technique is implemented through Eq. (2)

$$f_C(x) = \sum_{j=1}^{n} P_j(x) \frac{f(x) - \sum_{i \neq j} T_i P_i(x) \otimes h(x)}{P_j(x) \otimes h(x)} \quad (2)$$

where $f_C$ is the PVC corrected picture, $\otimes$ stands for the convolution operator, $h$ denotes the system's PSF, $f$ denotes the input (original) image, $P_{i,j}$ indicates the brain/anatomical regions, and $T_i$ is the corrected mean value of region *(i)* obtained from the GTM technique.

*B.3. Region-based voxel-wise correction (RBV):*

The RBV approach [28] is a hybrid methodology that employs the algorithm developed by Yang et al. [36] to accomplish both region- and voxel-based PVC. The mean activity levels in each region are first determined using the GTM approach. Afterward, Eq. (3) is employed to carry out a voxel-by-voxel PVC.

$$f_C(x) = f(x) \left[ \frac{s(x)}{s(x) \otimes h(x)} \right] \quad (3)$$

Here, $s(x) = \sum_{i=1}^{n} [T_i P_i(x)]$ stands for a synthetic image generated by the GTM technique (mean value is used for each region).

*B.4. Iterative Yang (IY):*

The IY methodology [29] is similar to Yang's method [36], wherein instead of applying the GTM method, the mean values of individual areas are calculated from the input PET data itself. The mean value estimations are then adjusted by applying Eq (4) in an iterative fashion.

$$f_{k+1}(x) = f(x)\left[\frac{s_k(x)}{s_k(x) \otimes h(x)}\right] \quad (4)$$

Here, $T_{k,i}$ is the estimated mean value of region *(i)* at the iteration *(k)*, and $f_k$ is an estimation of the PVE corrected image at the iteration *(k)*. In this equation, $s_k(x) = \sum_{i=1}^{n}[T_{k,i}P_i(x)]$ is a piece-wise version of the PET image with a mean value for each region.

*B.5. Deconvolution techniques*

Without any requirement for anatomical information or segmentation, two deconvolution techniques are applied to reduce the PVE. However, the efficiency of these procedures is restricted, and in some situations, these methods may result in noise amplification. When no anatomical information is provided, they can improve the contrast of the image to some extent. The iterative implementation of the two standard deconvolution algorithms, namely RL and RVC, is given in Eq 5 and Eq 6 [30].

$$f_{k+1}(x) = f_k(x)\left[h(x) \otimes \frac{f(x)}{h(x) \otimes f_k(x)}\right] \quad (5)$$

$$f_{k+1}(x) = f_k(x) + \alpha . h(x) \otimes [f(x) - h(x) \otimes f_k(x)] \quad (6)$$

Here, $\alpha$ is the convergence rate parameter that defines the degree of change at each iteration.

## C. Deep neural network implementation

A modified encoder-decoder U-Net architecture, illustrated in Figure 1, was employed to apply PVC to PET images. The modified U-Net is a fully convolutional neural network with deep concatenation connections between different stages, which is based on the encoder-decoder structure [37]. Convolution with 33 kernels and a corrected linear unit (ReLU) is used at each level. A 2×2 max-pooling is used for down-sampling and up-sampling blocks across various stages. The Adam optimizer was used with a learning rate of 0.003 to train the network. The input to the network was either 5% LD or FD PET images to predict partial volume corrected PET images. The goal was to train this network to perform the PVE correction on PET images based on the different aforementioned PVC approaches (independently). Considering the six PVC techniques, a total of 12 separate models (6 for LD and 6 for FD input data) were developed.

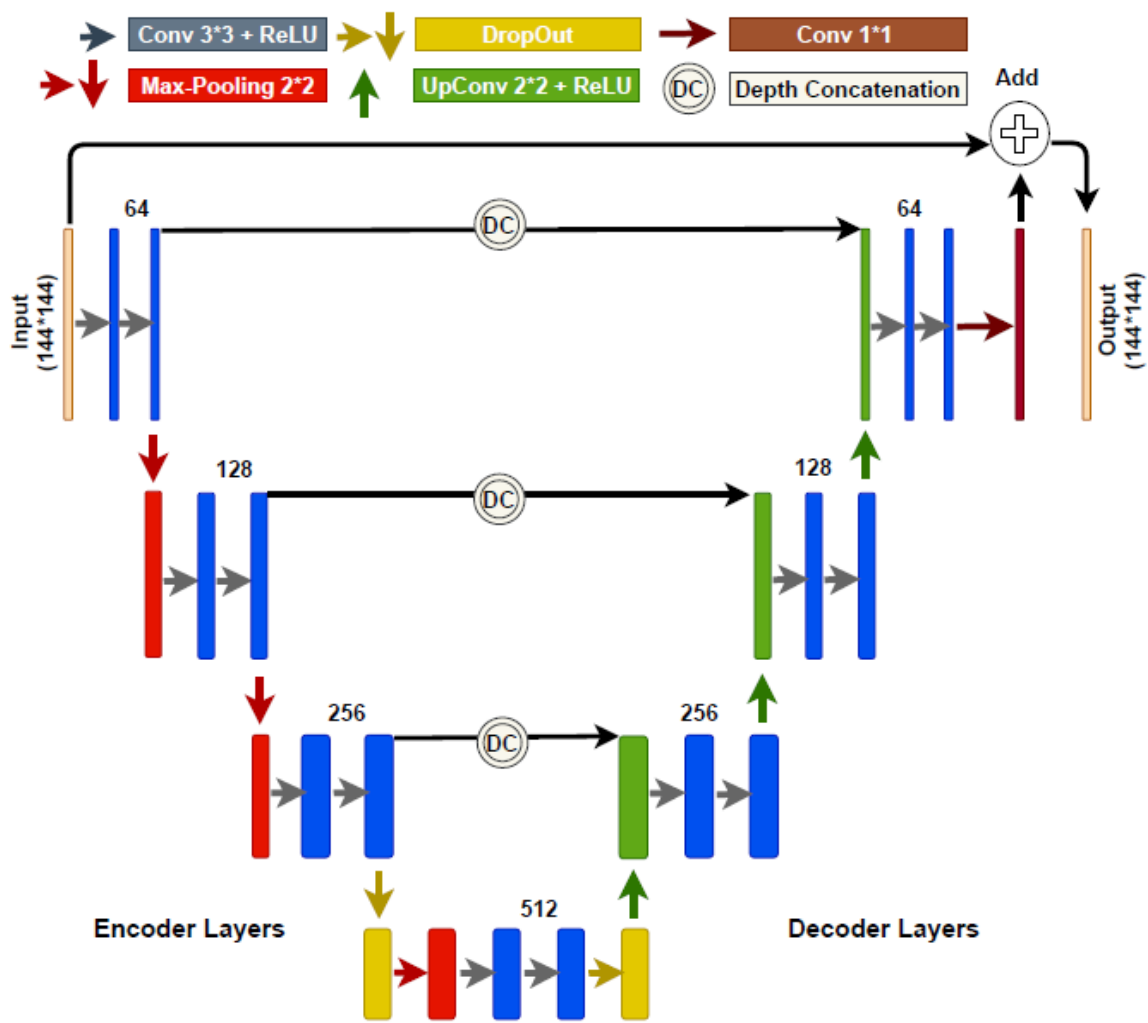

**Figure 1.** Structure of the U-Net employed for the prediction of partial volume corrected full-dose PET images from low-dose PET images (as well as full-dose PET images).

**D. Performance evaluation**

Three standard quantitative measures were performed to assess the quality of predicted PET pictures (FD+PVC) by different U-Net models. These metrics include SSI (Eq. 7), RMSE (Eq. 8), and PSNR (Eq. 9).

$$SSI(x,y) = \frac{(2\mu_x\mu_y + c_1)(2\sigma_{xy} + c_2)}{(\mu_x^2 + \mu_y^2 + c_1)(\sigma_x^2 + \sigma_y^2 + c_2)} \quad (7)$$

Here, $\mu_x, \mu_y$ represent the average of $x$, $y$ images and $\sigma_x^2, \sigma_y^2$ represent their variance. $\sigma_{xy}$ denotes the covariance of $x$ and $y$ images. $c_1 = (k_1 L)^2$, $c_2 = (k_2 L)^2$, k 1=0.01 and k 2=0.03 were selected by default to avoid division by very small numbers (L is the dynamic range of the pixel values).

$$RMSE = \sqrt{\frac{1}{M \times N} \sum_{i=1,j=1}^{i=M,j=N} (J(i,j) - I(i,j))^2} \quad (8)$$

Here, *J* denotes the anticipated PVC image, *I* is the reference image, and *M×N* denotes the number of pixels in the image.

$$PSNR = 20.\log_{10}\left(\frac{I_{max}}{RMSE}\right) \quad (9)$$

In Eq. (9), $I_{max}$ is the maximum value of either the reference or predicted PVC images.

For different PVC methods with either FD or LD input data, RMSE, PSNR, and SSI metrics were calculated and compared to the FD+PVC images predicted by the U-Net. Furthermore, given the AAL atlas transferred/co-registered to the patient's space (to define 71 anatomical brain areas), the region-wise RMSE was determined for the individual brain regions. The left and right regions were combined to reduce the number of brain areas to 34 for the evaluation of the models.

## III. Results

Figure 2 depicts representative LD and FD PET images (first row) together with the ground truth FD+PVC (second row) and FD+PVC images predicted by the U-Net models for different PVC algorithms (the third and fourth row). Visual inspection revealed that images predicted by the U-Net from both LD and FD PET images bear similar structures/details with respect to the reference FD+PVC images.

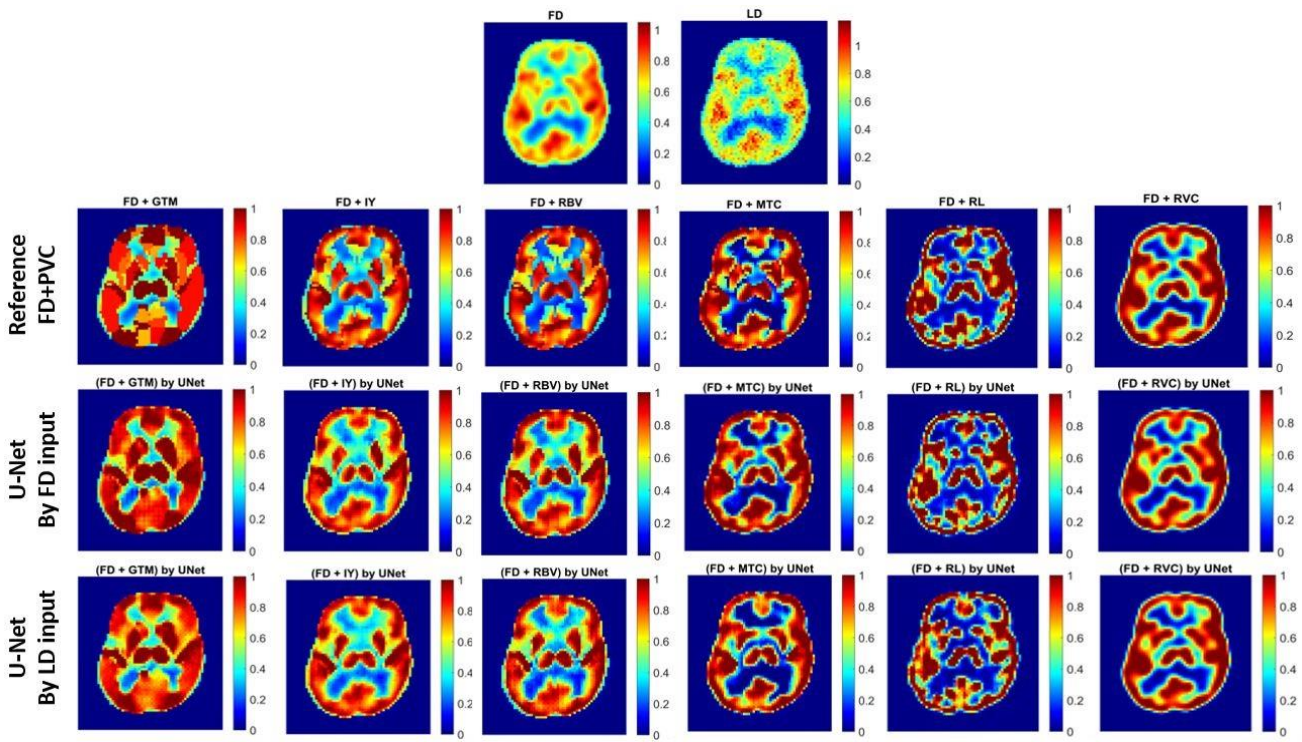

**Figure 2.** Representative LD and FD PET images before and after PVC. First row: full-dose and low-dose PET. Second row: reference FD+PVC PET images by different PVC methods. Third row: FD+PVC predicted by the U-Net models from FD PET input data. Fourth row: FD+PVC predicted by the U-Net models from LD PET input data.

Tables 1 and 2 illustrate the results of the quantitative analysis of the predicted PET images in terms of PSNR, RMSE, and SSI metrics for the LD and FD input PET data, respectively. These tables compare quantitative image quality metrics (mean±SD) between FD+PVC predicted by the U-Net models and the ground truth FD+PVC PET images for the entire head region. Among the PVC models, the RVC model exhibited better PSNR and RMSE values (3.6% and 11.25%, respectively), compared to the LD PET image. For the models with FD input data, similar to the LD input, slightly better results were obtained from the RVC model, and the worst results were from the RL model. Overall, the images predicted by both models with LD and FD PET input data exhibited comparable levels of error.

Table 1. Comparison of quantitative image quality metrics (mean±SD) calculated between FD+PVC predicted by the U-Net models and the ground truth FD+PVC PET images for the entire head region when the input data was LD PET images. The differences between LD and FD PET images are also presented.

|  | LD vs. FD | U-Net vs. FD+GTM | U-Net vs. FD+IY | U-Net vs. FD+RBV | U-Net vs. FD+MTC | U-Net vs. FD+RL | U-Net vs. FD+RVC |
|---|---|---|---|---|---|---|---|
| **PSNR** | 20.01±0.07 | 12.52±3.75 | 15.70±3.20 | 12.27±3.47 | 3.32±4.24 | 0.67±5.07 | 20.73±4.01 |
| **RMSE** | 0.89±0.16 | 1.65±0.66 | 1.30±0.47 | 1.60±0.61 | 2.71±1.41 | 5.00±2.39 | 0.99±0.45 |
| **SSI** | 0.54±0.09 | 0.63±0.05 | 0.63±0.06 | 0.57±0.06 | 0.64±0.05 | 0.30±0.07 | 0.89±0.05 |

Table 2. Comparison of quantitative image quality metrics (mean±SD) calculated between FD+PVC predicted by the U-Net models and the ground truth FD+PVC PET images for the entire head region when the input data was FD PET images.

|  | U-Net vs. FD+GTM | U-Net vs. FD+IY | U-Net vs. FD+RBV | U-Net vs. FD+MTC | U-Net vs. FD+RL | U-Net vs. FD+RVC |
|---|---|---|---|---|---|---|
| **PSNR** | 12.85±3.52 | 14.77±3.28 | 13.68±3.39 | 5.45±4.62 | 1.59±5.43 | 25.23±5.71 |
| **RMSE** | 1.64±0.62 | 1.35±0.48 | 1.48±0.55 | 2.45±1.34 | 3.27±1.71 | 0.79±0.48 |
| **SSI** | 0.29±0.05 | 0.67±0.06 | 0.65±0.05 | 0.73±0.05 | 0.87±0.04 | 0.94±0.05 |

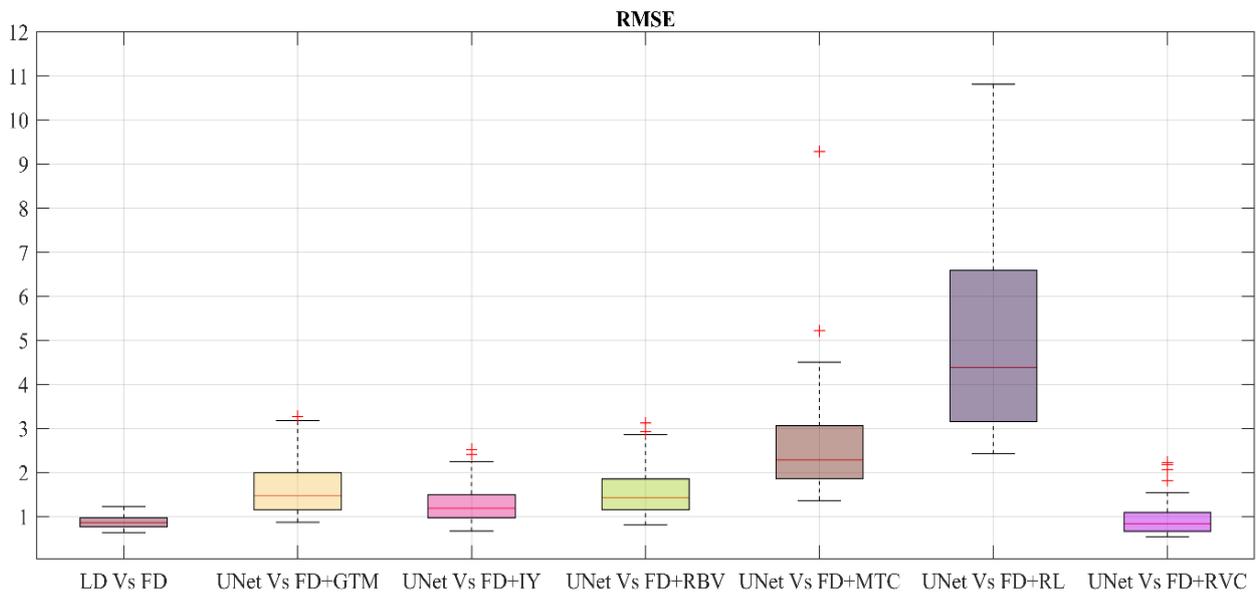

**Figure 3.** Boxplots of RMSEs for different PVC methods predicted by the U-Net models from LD PET input data. RMSEs between LD and FD PET images are also plotted.

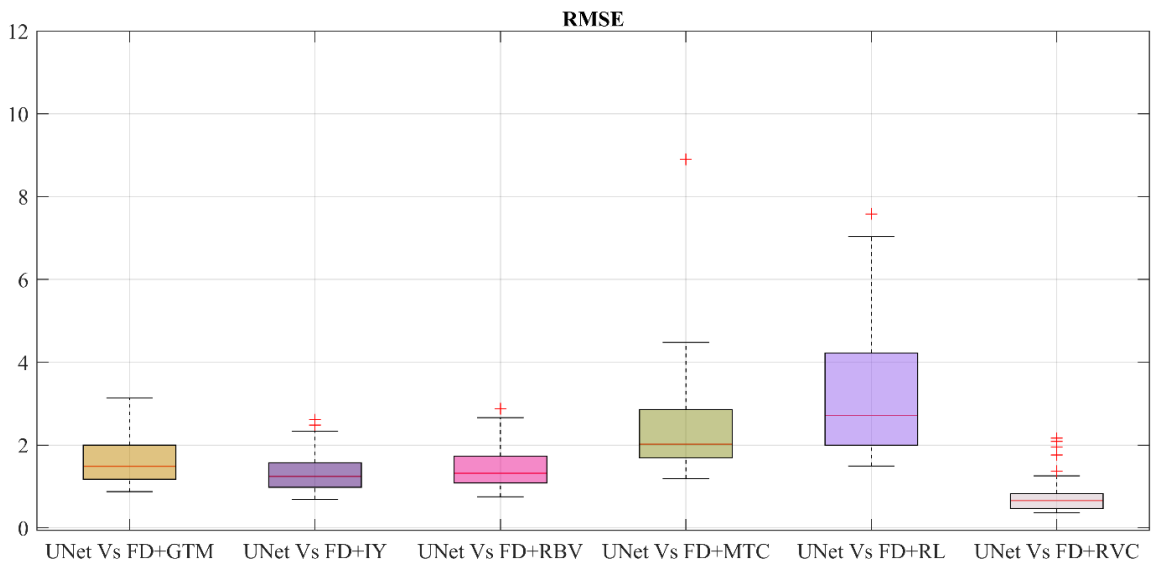

**Figure 4.** Boxplots of RMSEs for different PVC methods predicted by the U-Net models from FD PET input data.

**Error! Reference source not found.** and 4 show RMSE boxplots for the FD+PVC PET images predicted by different models from the LD and FD PET input data, respectively. The RMSE boxplots revealed that images predicted by the U-Net from LD and FD PET data contain similar levels of error, compared to the reference FD+PVC images.

The summary of the region-wise (for the different brain regions) analysis of different deep learning-based PVC models is presented in Figures 5 and 6 for LD and FD PET input data, respectively. Region-wise RMSEs were calculated using the AAL brain atlas to delineate 71 brain regions. The left and right regions were merged to reduce the number of brain regions to 34. For the models with LD input data, the RMSE values for predicted FD+GTM were higher in Heschl, Thalamus, Pallidum, Putamen, Caudate Nucleus, Amygdala, and Olfactory, compared to that in the LD PET images. For the predicted FD+IY, FD+RBV, and FD+MTC PET images, higher RMSEs were observed in Heschl, Thalamus, Pallidum, Putamen, and Caudate Nucleus brain regions, compared to LD PET images. For the RL model, higher RMSEs were observed in all brain regions. On the other hand, RVC showed better results in all brain regions, compared to the LD PET. Overall, the predicted partial volume corrected PET images from either LD or FD PET data exhibited similar results in the entire brain regions.

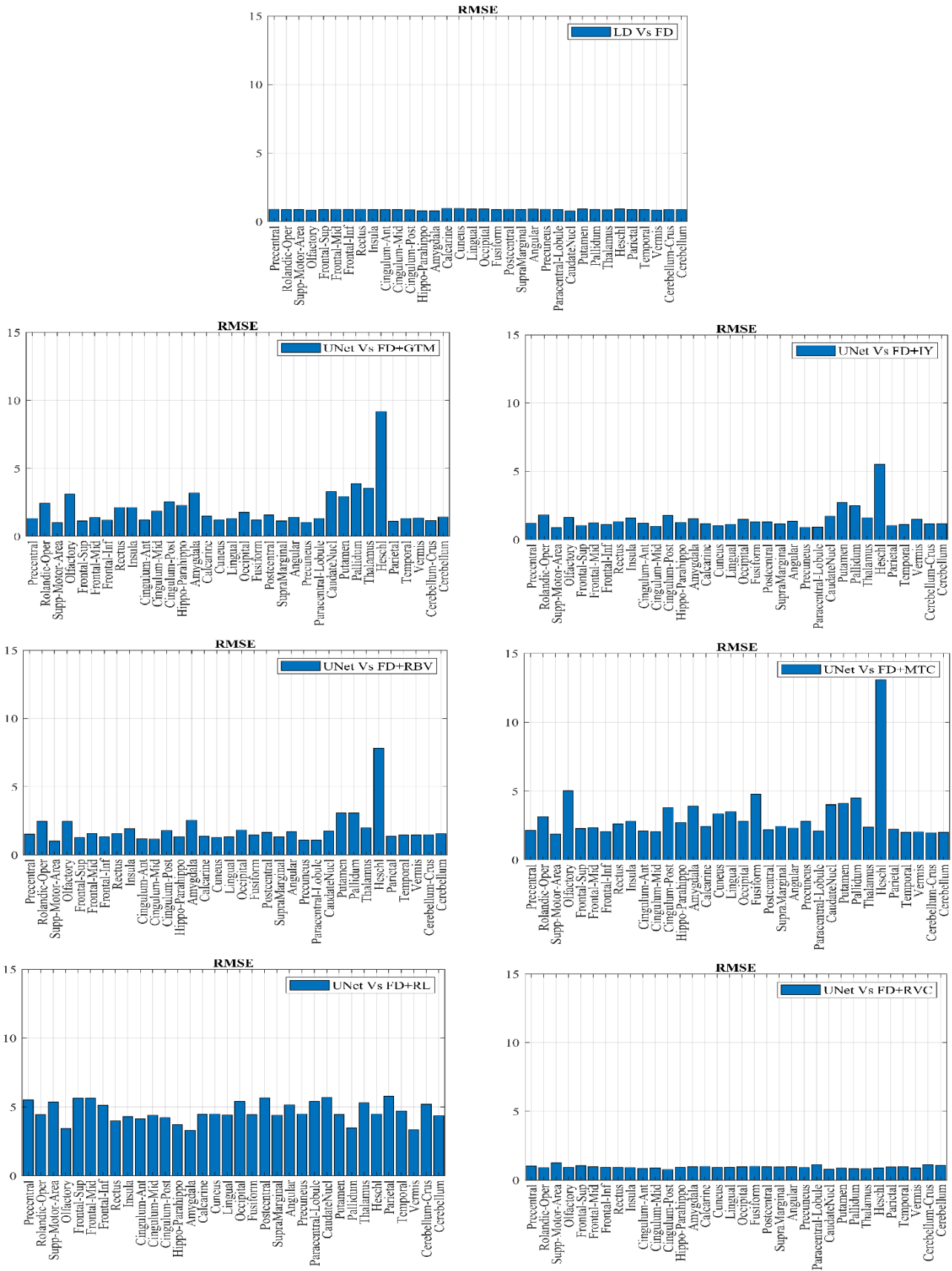

**Figure 5.** Bar plots of RMSEs calculated within different brain regions for different deep learning-based PVC models when the input data was LD PET images. Comparison between LD and FD PET images is also presented.

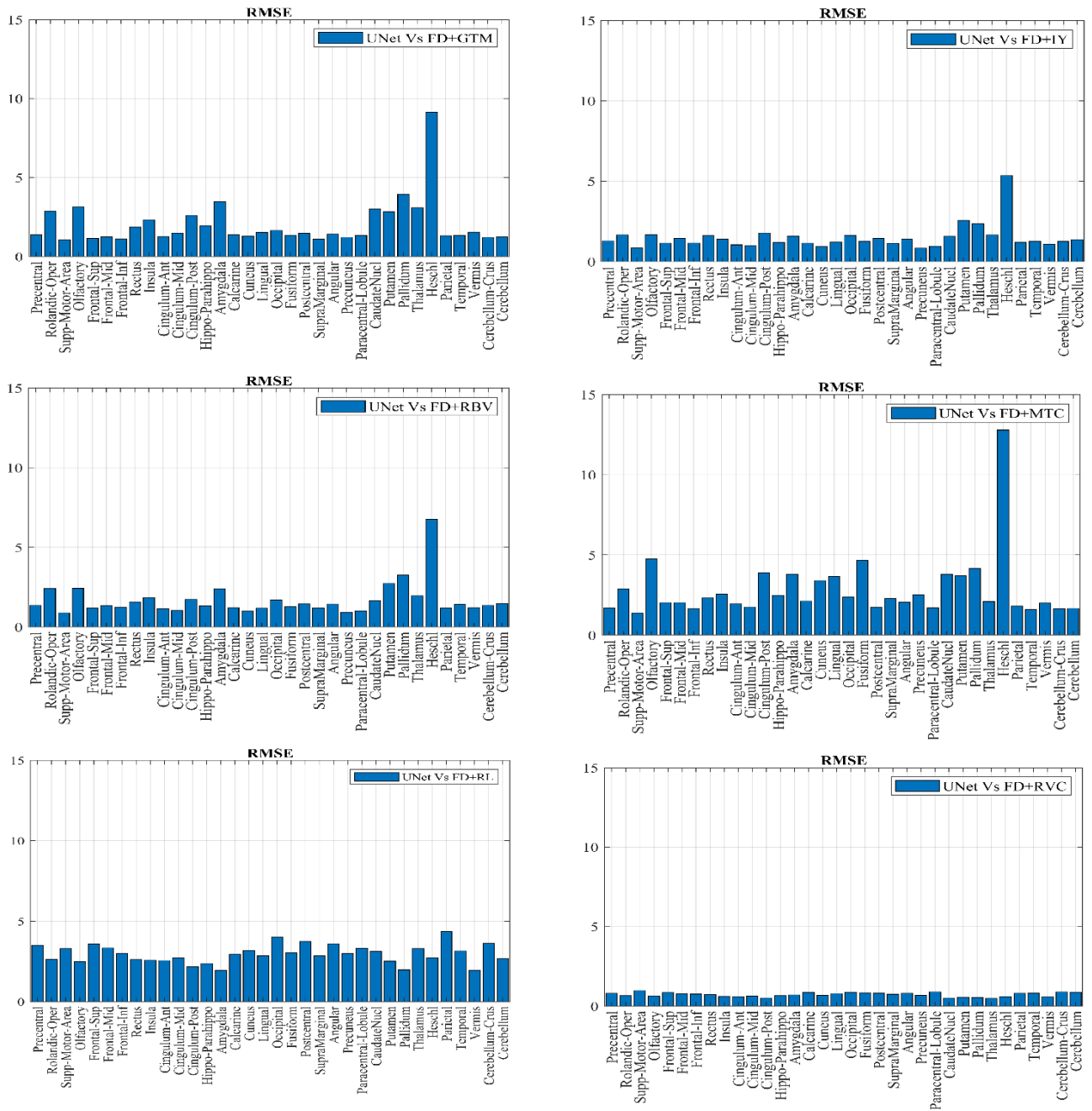

**Figure 6.** Bar plots of RMSEs calculated within different brain regions for different deep learning-based PVC models when the input data was FD PET images.

## IV. Discussion

We set out to create a deep learning solution for the joint PVC and denoising of LD PET data without using any anatomical pictures or information. This framework could be used in PET/CT imaging applications where MR pictures are not accessible for conventional PVC methods. To achieve this, we utilized FD and 5% LD PET data (as the input data) and FD PET pictures corrected for PVE using six frequently used PVC techniques to train a modified encoder-decoder U-Net network. To conduct PVC, the AAL brain atlas was used to delineate anatomical areas of the brain. The anticipated PVE corrected PET images from both LD and FD PET data revealed comparable image quality considering the visual assessment. The differences between LD and FD images (Table 1 and Figure 3) were significantly lower than those observed for PVC models since the errors due to the PVC did not exist in the comparison of the LD and FD images and it merely reflects the noise levels in the LD PET images. Considering the same levels of error observed in the models with LD and FD PET input data, it could be concluded that these models are able to perform joint PVC and noise reduction.

Since the RVC PVC strategy depends purely on a simple convolution procedure, which would not be challenging for the deep learning model to predict, the RVC model performed marginally better than the other PVC models. On the other hand, PVC techniques that rely on the brain areas mask to correct PET images exhibited greater quantification errors. Defining the exact anatomical brain regions binary would be highly challenging for the deep learning model, particularly when the input data is LD PET images. In this regard, significant RMSEs were observed for the MTC and RL PVC approaches, which could not be overlooked, and the current deep learning models would not be efficient for these two approaches.

The GTM approach is a region-based method that performs PVC concurrently on different brain regions provided by the AAL brain map. The GTM approach tends to estimate a mean value for each region, based on which the spill-in and spill-out across different regions are then calculated. Since this approach assumes a uniform activity uptake for each region, the accuracy of the PVC by the deep learning model solely depends on the identification of anatomical regions. However, the MTC approach is a hybrid PVC that relies on both region-based and voxel-level processing to correct the PVE. The MTC method, in the first step, creates an estimation of the corrected image based on which voxel-based processing is conducted to generate the final PVC image. This procedure is relatively more complicated, compared to the RVC approach, and thus, a larger RMSE was observed for this method. Similarly, the RBV PVC approach, which is a combination of Yang's voxel-wise correction [36] and the GTM approach, is a hybrid method; therefore, a similar RMSE to that of the MTC approach was observed. The RBV approach differs from the MTC method in that all areas are corrected at the same time while the MTC method performs PVC region by region [38, 39]. In the IY technique, which is a variation of the Yang's approach, the voxel-wise uptake values are taken directly from the input PET images, as opposed to the GTM method which relies on regional mean values. The IY is often faster than the RBV (and GTM) method/s to execute since it requires $k$ convolution operation, where $k$ is usually less than 10. However, the RBV method involves computing $n$ convolutions, where $n$ is the number of brain regions. Visually, the results of these two methods are similar. The PVC approaches discussed above are all based on anatomical information provided by the AAL brain map to calculate inter-region spill-over. Voxel-based PVC methods would also consider the intra-region spill-over and signal correction; thus, larger RMSEs were observed for these approaches.

Compared to the LD PET data, the predicted FD+GTM PET images in the Heschl, Thalamus, Pallidum, Putamen, Caudate Nucleos, Amygdala, and Olfactory regions showed an overestimation of activity. This observation could be due to fact that these regions have very close dimensions to the full width at half the maximum of the PSF of the system. In the predicted FD+IY, FD+RBV, and FD+MTC images, a larger over-estimation was observed in Heschl, Thalamus, Pallidum, Putamen, and Caudate Nucleos, in comparison with the LD PET images. These regions are of small size with relatively high activity levels, wherein small boundary estimation errors by the deep learning models would lead to large errors in these areas. Similarly, a larger overestimation was observed in Heschl, Thalamus, Pallidum, Putamen, and Caudate Nucleos regions for IY, MTC, and RBV models.

Since IY, RBV, and MTC techniques relied on a similar initial estimation generated by the GTM method, the models developed based on them exhibited comparable levels of error. For all brain regions, the RL model led to an overestimation of activity uptake, which might be attributable to the fact that this method does not rely on any initial PVC estimation. Overall, the RVC model exhibited lower errors in the entire brain regions, compared to the LD PET image for either LD or FD PET input data since this method contains a simple deconvolution process which is not challenging for deep learning models to predict.

One major limitation of this study is that it did not involve patient-specific anatomical MR data to correct the PVE, and instead, the AAL brain map was employed to apply PVC algorithms. Moreover, due to the lack of ground-truth data (such as simulation data) for the evaluation of the PVC algorithms, the performance of these techniques could not be compared in this study. In this light, some of these algorithms might have intrinsic limitations/shortcomings which may adversely affect the performance of the deep learning models developed based on them. To this end, there should be some standard approaches, such as the unrolling technique [40] so that we can decouple whether the inaccuracy is due to the intrinsic shortcoming of the PVC algorithm or the suboptimal deep learning development.

### V. Conclusion

Based on several PVC methods as the reference, we demonstrated that deep neural networks could conduct PVC on PET scans without anatomical images. Regarding the complexity of PVC techniques, varying levels of error were observed. The deep learning-based PVC models exhibited similar performance for both FD and LD input PET data, which demonstrated that these models could perform joint PVC and noise reduction. Since these models do not require any anatomical images, such as MR data, they could be used in dedicated brain PET, as well as PET/CT scanners.